\documentclass[trackchanges]{aastex701}

\usepackage{hyperref}
\usepackage{amsmath}
\usepackage{enumerate}
\usepackage[utf8]{inputenc}
\usepackage{cleveref}
\usepackage{amssymb}
\usepackage[T1]{fontenc}
\usepackage{enumitem}
\usepackage{url}
\usepackage[misc]{ifsym}
\usepackage{multirow}
\usepackage{mhchem}
\usepackage{booktabs}
\usepackage{CJKutf8}
\usepackage{subfigure}

\begin{document}

\title{Periodic Radio Technosignature Search toward 3I/ATLAS with FAST}

\begin{CJK*}{UTF8}{gbsn}
\author[0000-0002-1190-473X]{Jian-Kang Li (李健康)}
\affiliation{School of Integrated Circuits, Shanghai Dianji University, Shanghai 200240, People's Republic of China}
\affiliation{Institute for Frontiers in Astronomy and Astrophysics, Beijing Normal University, Beijing 102206, People's Republic of China}
 \affiliation{School of Physics and Astronomy, Beijing Normal University, Beijing 100875, People's Republic of China}
 \email{fakeemail1@google.com}  

\author[0000-0002-4683-5500]{Zhen-Zhao Tao (陶振钊)}
\affiliation{College of Computer and Information, Dezhou University, Dezhou 253023, People's Republic of China}
\email{fakeemail1@google.com}

\author[0000-0002-3363-9965]{Tong-Jie Zhang (张同杰) \href{mailto:tjzhang@bnu.edu.cn}{\textrm{\Letter}}}
\affiliation{Institute for Frontiers in Astronomy and Astrophysics, Beijing Normal University, Beijing 102206, People's Republic of China}
\affiliation{School of Physics and Astronomy, Beijing Normal University, Beijing 100875, People's Republic of China}
\email[show]{tjzhang@bnu.edu.cn} 

\author{Men-Quan Liu (刘门全)\href{mailto:menquan@sdju.edu.cn}{\textrm{\Letter}}}
\affiliation{School of Integrated Circuits, Shanghai Dianji University, Shanghai 200240, People's Republic of China}
\email[show]{menquan@sdju.edu.cn}

\begin{abstract}
3I/ATLAS, the third confirmed interstellar object discovered in the Solar System, provides a unique opportunity for targeted technosignature searches. We report a periodic radio technosignature search toward 3I/ATLAS using the Five-hundred-meter Aperture Spherical Telescope (FAST) L-band multibeam receiver. To search for periodically modulated signals and distinguish center-beam-dominated candidates from multibeam radio-frequency interference, we apply canonical polyadic decomposition (CPD) to the multibeam dynamic spectra. CPD factorizes the multibeam data tensor into a set of separable components, with associated time, frequency, and beam signatures. Candidate components are then selected through periodogram and autocorrelation diagnostics. We find no credible artificial periodic radio technosignature above 0.146 W is detected from the direction of 3I/ATLAS. This search expands the range of signal types explored for this target by including periodic modulated signal, and illustrates that CPD is a promising framework for multibeam periodic technosignature searches.

\end{abstract}

\keywords{\href{http://astrothesaurus.org/uat/2127}{Search for extraterrestrial intelligence (2127)}; \href{http://astrothesaurus.org/uat/1338}{Radio astronomy (1338)};\href{http://astrothesaurus.org/uat/52}{Interstellar objects(52)}; \href{http://astrothesaurus.org/uat/74}{Astrobiology (74)}}


\section{Introduction}\label{sec:introduction}

The Search for Extraterrestrial Intelligence (SETI) aims at identifying technosignatures as observable manifestations of technology \citep{2001ARA&A..39..511T}. Among the various proposed technosignatures, radio signals remain one of the most extensively explored channels because of their detectability, transmission efficiency, and long propagation range \citep{1959Natur.184..844C}. Beyond the conventional focuses on narrowband Doppler-drifting signals, SETI searches can also be extended to other phenomenological classes of radio technosignatures. Periodically modulated radio signals constitute a plausible technosignature class because artificial transmitters need not remain continuously on, and may instead appear as intermittent or repeating signals owing to beacon operation, sky-scanning strategies, rotational or orbital modulation, or energy cost consideration. Cost-optimized beacon models, for example, have explicitly argued that economically motivated interstellar transmitters may be observed as recurrent short-duration events rather than as continuous emission\citep{2010AsBio..10..491B}, while intermittent signaling has also been discussed as a natural consequence of planetary rotation, duty-cycle constraints, and targeted transmission geometry \citep{2020IJAsB..19..299G}. In parallel with these theoretical considerations, periodic SETI works have already been explored observationally in several in radio observations\citep[e.g.][]{2018ApJ...869...66H,2023AJ....165..255S,2026AJ....171...78L}.

3I/ATLAS is the third confirmed interstellar object passing through the Solar System. As a rare extrasolar body observed at close range during a limited time window, it provides an unusual target for sensitive radio technosignature searches. Dedicated SETI observations of 3I/ATLAS have already been reported with several facilities, including the Allen Telescope Array, MeerKAT, and the Green Bank Telescope, with no credible narrowband drifting technosignature detections reported to date \citep{2026AJ....172....1S,2025ATel17499....1P,2025RNAAS...9..351J}. FAST, with its exceptional sensitivity and multibeam capability, offers a particularly powerful dataset for extending such searches \citep{2019SCPMA..6259502J,2020RAA....20...64J,2020Innov...100053Q}. We also carry out a narrowband Doppler-drift singal search using the same FAST multibeam observation data of 3I/ATLAS\citep{2026ApJ..1006..163L}.

Different from our previous narrowband drifting technosignature search, the present paper focuses on periodically modulated signal structures in the time. Relative to drifting narrowband searches, extending the accessible technosignature parameter space. This work adopts canonical polyadic decomposition \citep[CPD;][]{sapm192761164,carroll1970analysis,doi:10.1137/07070111X} as the central analysis framework for periodic signal identification in FAST multibeam dynamic spectra. By constructing the data tensor cube into time, frequency, and beam components, the method enables simultaneous characterization of temporal periodicity, spectral structure, and beam occupancy. In this formulation, periodic signal searching and center-beam discrimination are embedded in a unified decomposition framework, which is particularly suitable for multibeam SETI observations.

The focus of the present paper is placed on the periodicity search methodology and its application to the 3I/ATLAS observations. Section~\ref{sec:method} presents the CPD-based periodic-search procedure. Section~\ref{sec:Results} gives the search results and diagnostic characterization. The implications and limitations of the analysis are discussed in Section~\ref{sec:Discussion}, and Section~\ref{sec:Conclusion} summarizes the main conclusions.

\section{Methodology and Data Analysis}\label{sec:method}

The data used in this work was recorded by the FAST psr backend, configured with a sampling time of 49.152 $\mu{\rm s}$ and a frequency resolution of $\sim$ 0.122 MHz. The observation dates\footnote{Since we did not turn on the psr data backend in 2025-10-03, we only carry out periodic technosignature search for these three days in this work.} and observation modes have been introduced in \cite{2026ApJ..1006..163L}. During each observation, a 2 s reference diode signal was injected every 300 s for temperature and polarization calibration. This reference diode signal can also serve as an internal reference for validating the sensitivity of the periodic-signal search and for interpreting components associated with the calibration cadence. In addition, radio frequency interference (RFI) mitigation was performed using a root-mean-square (rms)-based flagging scheme, in which any frequency channel whose rms deviated from the median by more than $5\times{\rm MAD}$ was flagged. The polarization calibration and rms RFI flagged schemes are adopted from \cite{2021RAA....21..282S}.

For each observation, the time-frequency data from multiple beams are organized into a three-dimensional data cube and analyzed independently. In practice, the central beam still serves as the center beam, while only three of the outermost beams are selected as reference off-source beams. Additionally, the dynamic spectra used here have been time-binned, with a post-downsampling time bin of $\sim0.12~{\rm s}$, which allows an efficient search over moderate and relatively long periodic timescales while preserving the multibeam structure required for signal identification.\footnote{These choices were motivated by the very large data volume of the original multibeam data and the associated computational cost. In the present analysis, we therefore used only three outermost beams as reference beams and adopted time-downsampled data products. Future work may extend the analysis to a larger set of reference beams and explore analyses based on data with higher time resolution.} The analysis procedure consists of four main steps: construction of the multibeam data tensor with baseline subtraction and normalization preprocessing, CPD, identification of periodic components and multibeam comparison. In addition, an autocorrelation function (ACF) diagnostic is used to provide supplementary characterization of the frequency-domain structure of each component.

\subsection{Canonical Polyadic Decomposition and Beam-domain Characterization}

For each observation, the multibeam dynamic spectrum is represented as a third-order tensor
\begin{equation}
X(t,\nu,b)\in\mathbb{R}^{T\times F\times B},
\end{equation}
where $t$ denotes time index, $\nu$ denotes frequency channel, and $b$ denotes beam index. Here $T$, $F$, and $B$ are the numbers of time samples, frequency channels, and beams, respectively. This tensor representation preserves the three observables most relevant to the present search: temporal modulation, spectral localization, and beam-to-beam variation.

To separate mixed structures in the multibeam dynamic spectrum, we apply canonical polyadic decomposition (CPD), also known as the CANDECOMP/PARAFAC decomposition, to the tensor $X(t,\nu,b)$. CPD represents a tensor as a sum of rank-one outer products, with early roots in the polyadic tensor decomposition introduced by \citet{sapm192761164} and broad modern applications in signal processing and machine learning \citep[e.g.,][]{CONG201559,2017ITSP...65.3551S}. In astronomy, \citet{2018A&C....25..195F} introduced TensorFit as a tensor-mode framework for handling and analyzing large spectral cubes, motivated by data products from facilities such as ALMA. The spectral cube is treated as a multidimension data object, typically with two spatial axes and one spectral/velocity axis to support scalable cube analysis.

In our work, the multibeam dynamic spectra data can be regarded as a native tensor form. The data can be decomposed as
\begin{equation}
X_{t\nu b}\approx \sum_{r=1}^{R}\lambda_r\,\mathbf{a}_r \circ \mathbf{c}_r \circ \mathbf{d}_r =\sum_{r=1}^{R}\lambda_r\,a_{tr}\,c_{\nu r}\,d_{br},
\end{equation}
where $R$ is the adopted decomposition rank, $\lambda_r$ is the weight of the $r$th component, $a_{tr}$ is its temporal factor, $c_{\nu r}$ is its frequency factor, and $d_{br}$ is its beam factor. In this form, each component provides a separable description of one structure in the data across time, frequency, and beam space. Unlike matrix decomposition, however, the determination of tensor rank is substantially more complicated, and in general, has been proven as an NP-hard problem \citep{10.1145/2512329}, so in practice $R$ must be specified empirically. For a third-order tensor $X\in\mathbb{R}^{T\times F\times B}$, the rank is weakly bounded by \citet{10.5555/120565.120567}
\begin{equation}
R \leqslant \min(TF,\,TB,\,FB).
\end{equation}
In the present application, we restricted $R$ to a dimensionally admissible range and, in the final analysis, adopted $R = F$, i.e., the number of frequency channels. In this form, each component is described by a separable structure in time, frequency, and beam space. This choice provides sufficient component capacity for channel-scale structures while keeping the dense tensor decomposition computationally tractable. Details of the CPD implementation are described in Appendix \ref{app:cpals}.

The relative importance of each component is measured by its normalized energy contribution,
\begin{equation}
\eta_r=\frac{\lambda_r^2}{\sum_k \lambda_k^2},
\end{equation}
which is used to rank the energy level of components. In the following, the component index $r$ refers to the original decomposition label, while the energy rank refers to the ordering induced by $\eta_r$. To quantify the beam-domain concentration of each component, we define the beam fraction
\begin{equation}
f_{br}=\frac{\lambda_r^2 d_{br}^2}{\sum_{b'} \lambda_r^2 d_{b'r}^2},
\end{equation}
which measures the fractional beam-domain energy carried by beam $b$ for component $r$. For the center beam $b_c$, we further define a dominance ratio
\begin{equation}
D_r=\frac{f_{b_c r}}{\max_{b\neq b_c} f_{br}},
\end{equation}
so that larger $D_r$ indicates a stronger concentration toward the center beam relative to all off-center beams. We also compute the normalized beam entropy
\begin{equation}
H_r=-\frac{1}{\ln B}\sum_b f_{br}\ln f_{br},
\end{equation}
which measures the concentration of the component energy distribution across the beam axis of the $X_{t \nu b}$ data cube. A component with a large $D_r$ and a small $H_r$ is therefore considered more center-dominant in beam space. These quantities are used as practical diagnostics to identify components whose beam signatures are more consistent with center-beam localization.

\subsection{Periodicity Diagnostics and Multibeam Validation}

For each CPD component, we firstly examine the periodicity from the temporal factor through the periodogram. The minimum period is set to $P_{\min}=6\,t_{\rm samp}$, where $t_{\rm samp}$ is the effective sampling time of the time-binned data. This lower bound is adopted to ensure that even the shortest searched periodicity is resolved by multiple time samples. The maximum period is set to $P_{\max}=T_{\rm obs}/3$, where $T_{\rm obs}$ is the observing duration. This choice requires that a candidate periodic signal be sampled over at least about three cycles within a given observation.     For the present observations, this corresponds to $P_{\min}\simeq 0.7\,\mathrm{s}$ and $P_{\max}\simeq 10^3\,\mathrm{s}$. Additionally, the duty cycle $\delta=\tau_{\rm sig}/P$ search range is 0.01--0.5 in this work, where the $\tau_{\rm sig}$ is the signal duration of a single pulse.

The multibeam validation is based on a set of criteria designed to identify CPD components whose energy and periodicity are preferentially concentrated in the central beam. We define a per-beam temporal projection
\begin{equation}
Z^{(r)}_{t,b} \equiv \sum_\nu X(t,\nu,b)\,\tilde c_r(\nu), \qquad 
\tilde c_r(\nu)\equiv \frac{c_r(\nu)}{\|c_r\|_2},
\label{eq:Ztb_def}
\end{equation}
which yields a time series in each beam that is maximally sensitive to the component's spectral footprint. 
Periodicity is quantified from $Z^{(r)}_{t,b}$ via the periodogram, where the component period $P_0$ is obtained from the temporal factor $a_r(t)$. 
For consistency, we also measure the peak period in the center-beam projection, $P_{0,\rm center}$, and require it agreement with the component period $P_0$. We also set a minimum periodogram SNR at $P_0$ in the center beam together with an upper limit on the maximum off-beam periodogram SNR at the same $P_0$, so that only periodicities that are both significant and preferentially concentrated in the center beam are retained for further inspection. The beam dominance ratio $D_r$ and normalized beam entropy $H_r$ are used to measure the energy concentration in beam space for components. The adopted quantities and criteria are summarized in Table~\ref{tab:periodicity_multibeam_criteria}.

\begin{deluxetable}{ll}
\tablecaption{Adopted criteria for periodicity diagnostics and multibeam validation.\label{tab:periodicity_multibeam_criteria}}
\tablehead{
\colhead{Quantity} & \colhead{Criteria}
}
\startdata
Beam dominance ratio $D_r$ & $D_r \geqslant  3.0$ \\
Normalized beam entropy $H_r$ & $H_r \leqslant  0.60$ \\
Period consistency &
$\left|\log_{10} P_{0,\rm center} - \log_{10} P_{0}\right|\le 0.02$ \\
Central-beam periodogram SNR at $P_0$ & ${\rm SNR}_{\rm center}(P_0) \geqslant 10$ \\
Maximum off-beam periodogram SNR at $P_0$ & ${\rm SNR}_{\rm off,max}(P_0) \leqslant 8$ \\
\enddata
\end{deluxetable}

\begin{figure}[htbp]
  \centering
  \gridline{
    \fig{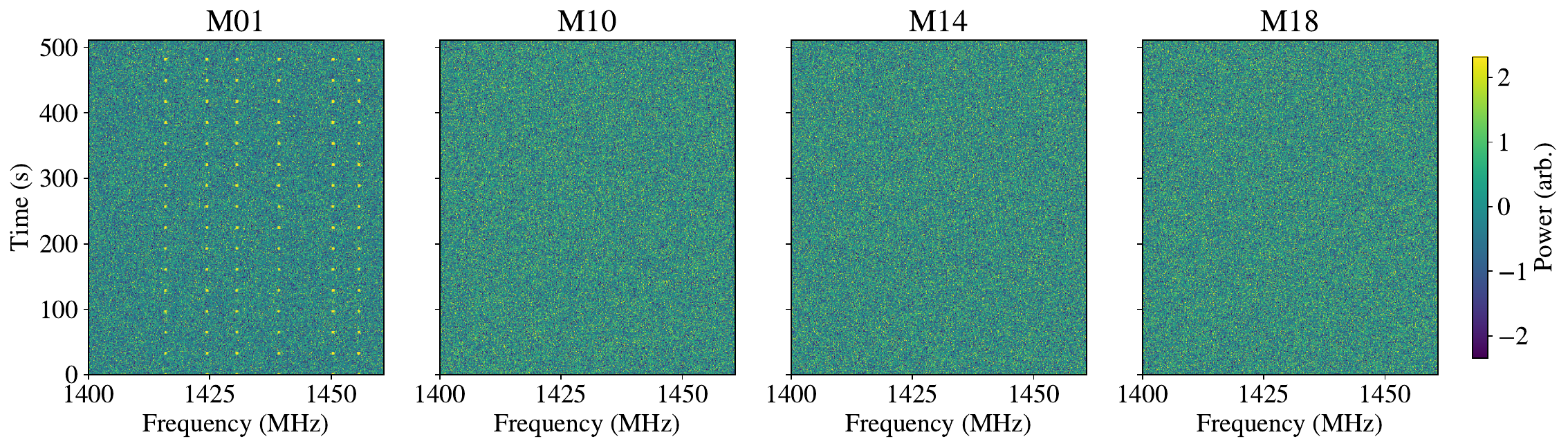}{0.8\textwidth}{(a)}
  }
  \gridline{
    \fig{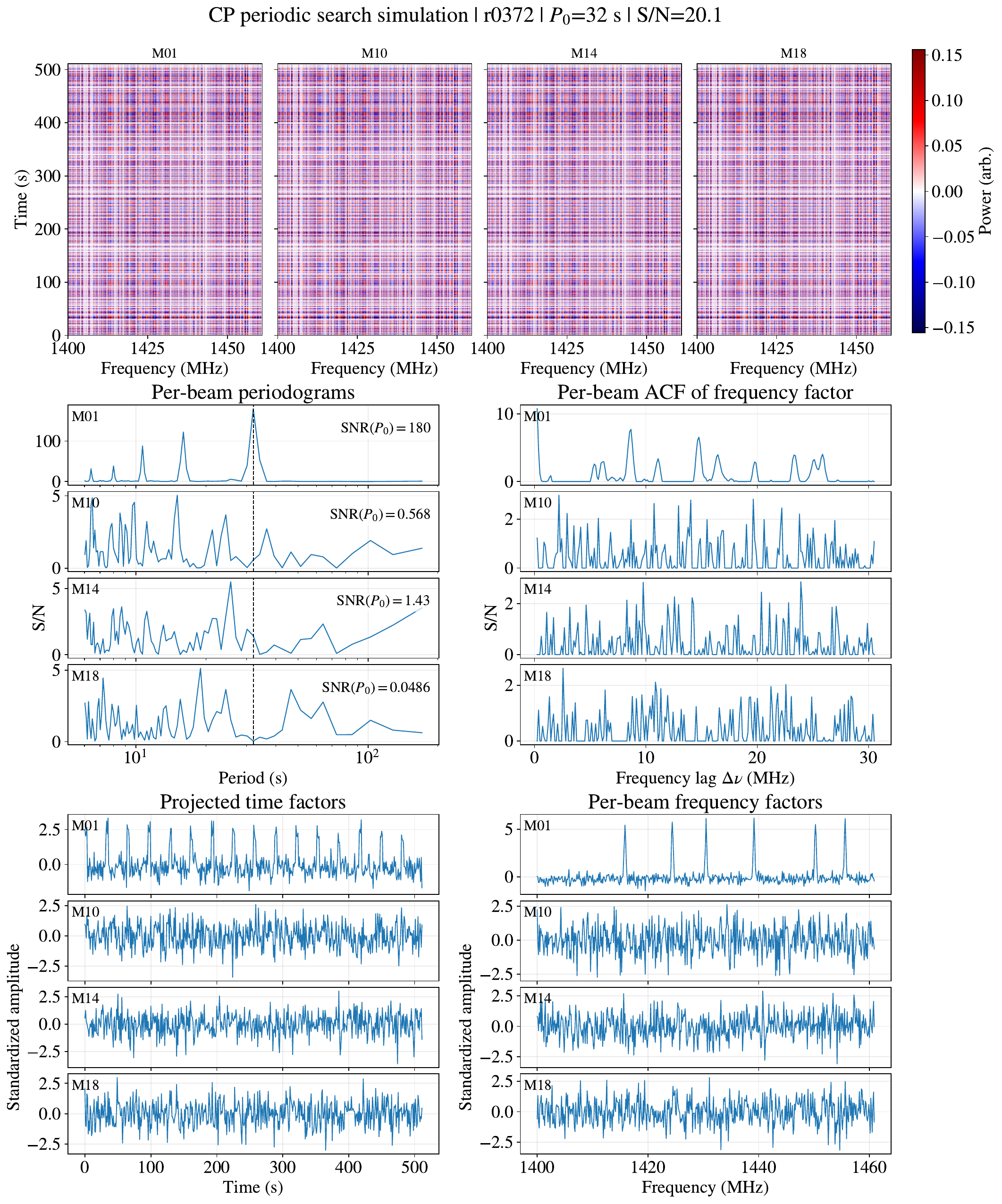}{0.7\textwidth}{(b)}
  }
    \caption{
    Simulated injection and CPD recovery of a center-dominant periodic signal.
    (a) Raw synthetic multibeam dynamic spectra before CPD. A periodic signal with period $P=32~{\rm s}$, duty cycle $\delta=0.12$ and peak amplitude $A=8\sigma_{\rm noise}$ is injected into six narrow frequency channels in the central beam only.
    (b) CPD-based diagnostics for the corresponding center-dominant component. Top row: the dynamic spectra for the four beam during the observation. Middle-left: periodograms of the per-beam projected time factors, with the component-defined period $P_0$ (vertical dashed line) and the corresponding periodogram significance $\mathrm{SNR}_b(P_0)$ annotated in each panel. Middle-right: per-beam ACF computed from the per-beam frequency projections. Bottom-left: normalized projected time series. Bottom-right: normalized per-beam frequency projections.}
    
    \label{fig:simulate_signal}
\end{figure}

In addition, we also define a per-beam spectral projection,
\begin{equation}
W^{(r)}_{\nu,b} \equiv \sum_t X(t,\nu,b)\,\tilde a_r(t), \qquad 
\tilde a_r(t)\equiv \frac{a_r(t)}{\|a_r\|_2},
\label{eq:Wfb_def}
\end{equation}
which highlights the frequency channels that participate in the component's temporal modulation in each beam. We compute the ACF of $W^{(r)}_{\nu,b}$ to characterize the spectral morphology associated with the components. In particular, channel-confined structures are expected to produce localized peaks in $W^{(r)}_{\nu,b}$, whereas some forms of RFI tend to exhibit comb-like features that leave a more organized equal-spaced peaks imprint in the ACF. Taken together, components that satisfy the beam-space concentration criteria ($D_r$ and $H_r$) and the multibeam periodicity criteria are classified as potential candidate components. Figure~\ref{fig:simulate_signal} shows a simulated periodic signal with $P_0=32$ s injected in the center beam only. After CPD, the injected signal is decomposed as a component whose periodicity is strongly supported in the central-beam projection but is not comparably present in the off-beam projections, yielding a large contrast between the periodograms. The corresponding frequency-domain projection is concentrated at the injected channels, producing several prominent isolated peaks. The ACF of frequency projection also correspondingly exhibits elevated values at non-zero lags associated with separations between these peaks.

\subsection{Steps for the Analysis Framework}

Before CPD analysis, the raw psrfits data are converted into filterbank format for four polarization channels, and thus derived the Stokes parameters as $I=XX+YY$, $Q=XX-YY$, $U=2XY$, $V=-2XY$ after polarization calibration. The multibeam dynamic spectra are assembled into a tensor for each Stokes product, CPD is then applied to the preprocessed multibeam data tensor. Our goal is not merely to identify components with periodicity in time, but also to characterize whether such periodicity is preferentially associated with the center beam rather than being comparably present in reference beams.

The analysis consists of two main parts. The first part is the CPD of each multibeam tensor, from which we record the basic properties of all components, including their component weights, energy ratios, beam-domain diagnostics, dominant periods, and auxiliary ACF-based frequency descriptors. This step provides a global characterization of the decomposed structures and serves as the basis for identifying components of potential interest.

The second is a component-level diagnostic product produced for the components. We further examine the temporal periodicity, compare the corresponding periodogram behavior across all available beams, and inspect the frequency-domain morphology through the ACF of the frequency factor. The per-beam periodic metrics and related diagnostic quantities are retained for subsequent visual inspection and statistical analysis.

\section{Results} \label{sec:Results}

After applying the RFI-flagging procedure described in Section~\ref{sec:method}, we identified a heavily contaminated frequency interval spanning approximately 1140--1300~MHz, where the interference level was markedly higher than that in the rest of the band. This interval was therefore masked in all subsequent processing. The analysis was then restricted to the two retained sub-bands, 1050--1140~MHz and 1300--1450~MHz, and the total component number for each data tensor is fixed to $N_{\rm comp}=1965$. 

Overall, we find 2753 central beam dominant components ($D_r\geqslant  3.0$ and $H_r \leqslant  0.60$) for this periodic signals search toward 3I/ATLAS. Among these central beam dominant components, we find 3 components satisfy the criteria in Table~\ref{tab:periodicity_multibeam_criteria}, and their information is summarized in Table~\ref{tab:cand_info}. These three components were all found in Stokes-$V$ polarization channel during the observation on 2026-01-05. The component r1176 is rejected as part of the injected calibration signal due to its close agreement with the calibration timescale. The component r55 exhibits periodogram peak at $P_0=18.874$ s in central beam, but the periodogram shows a remarkable harmonic structure and also retains substantial power $\sim$300 s, and the component time factor displays a clear modulation on $\sim$300 s timescale, we therefore attribute r55 as another manifestation of the calibration injection rather than an independent narrowband periodic signal. Component r952 does not exhibit the characteristic $\sim$300 s harmonic pattern. However, the central beam frequency factors do not show a clear localized spectral concentration, and the corresponding frequency-domain ACFs lack a distinct, reproducible correlation scale beyond weak fluctuations. We thus regard r952 as inconsistent with the narrowband periodic signal we search and exclude it.
\begin{deluxetable}{lccccccc}
\tablecaption{Information of the potential candidate components.\label{tab:cand_info}}
\tablehead{
\colhead{Components} & \colhead{Energy rank order} & \colhead{$D_r$} & \colhead{$H_r$} & \colhead{$P_0$ (s)} & \colhead{$P_{0,\rm center}$ (s)} & \colhead{${\rm SNR}_{\rm center}(P_0)$} & \colhead{${\rm SNR}_{\rm off,max}(P_0)$} 
}
\startdata
r1176 & 793 & 11.683 & 0.340 & 300.097 & 300.097 & 1025.542 & 6.240 \\
r55   & 315 & 23.039 & 0.144 & 18.874 & 18.874 & 	701.013 & 6.188 \\
r952  & 1212& 56.775 & 0.072 & 16.580 & 16.580 & 13.359 & 3.388 \\
\enddata
\end{deluxetable}

\begin{figure}[htpb]
  \centering
  \includegraphics[width=0.495\textwidth]{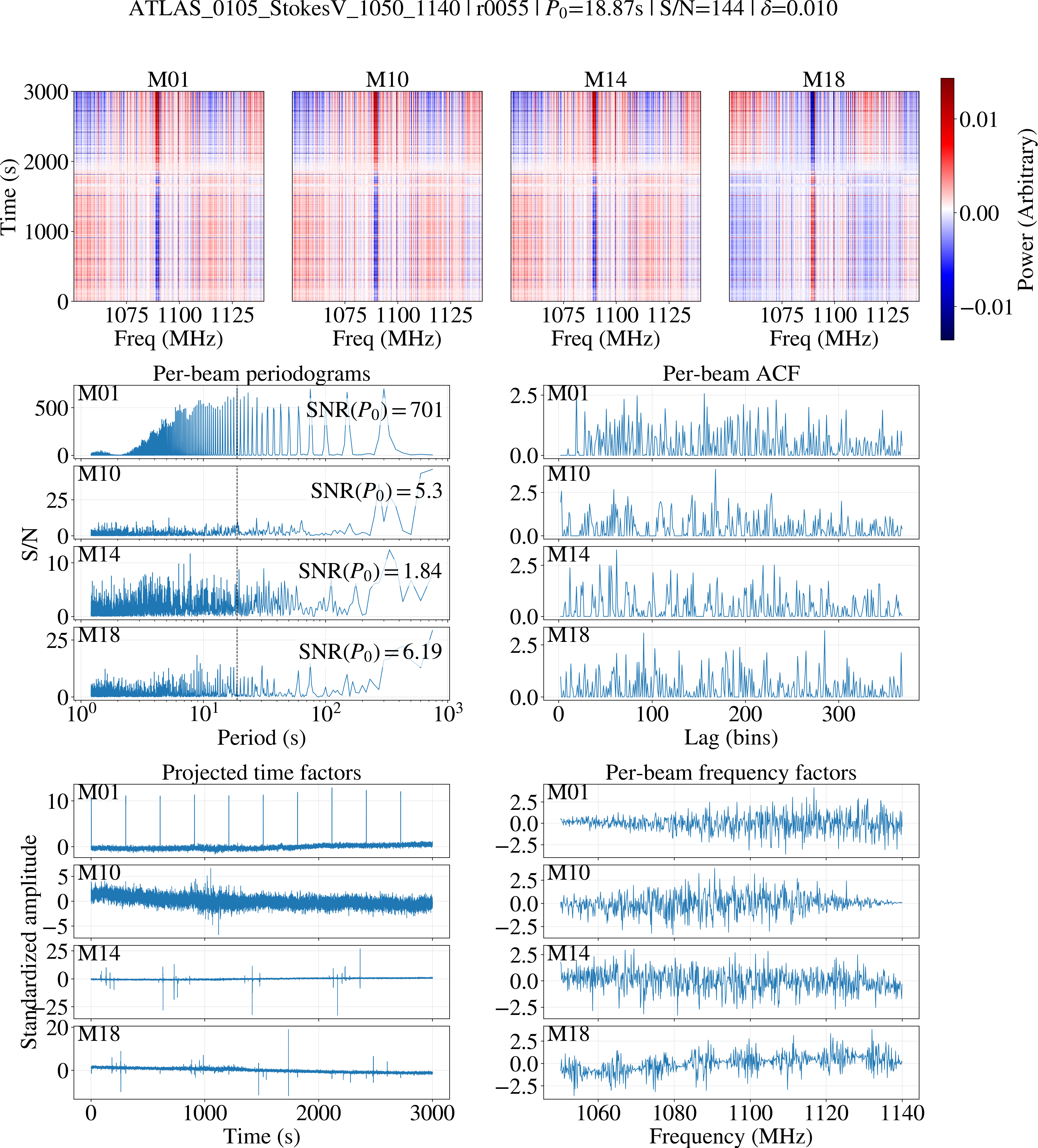}
  \includegraphics[width=0.495\textwidth]{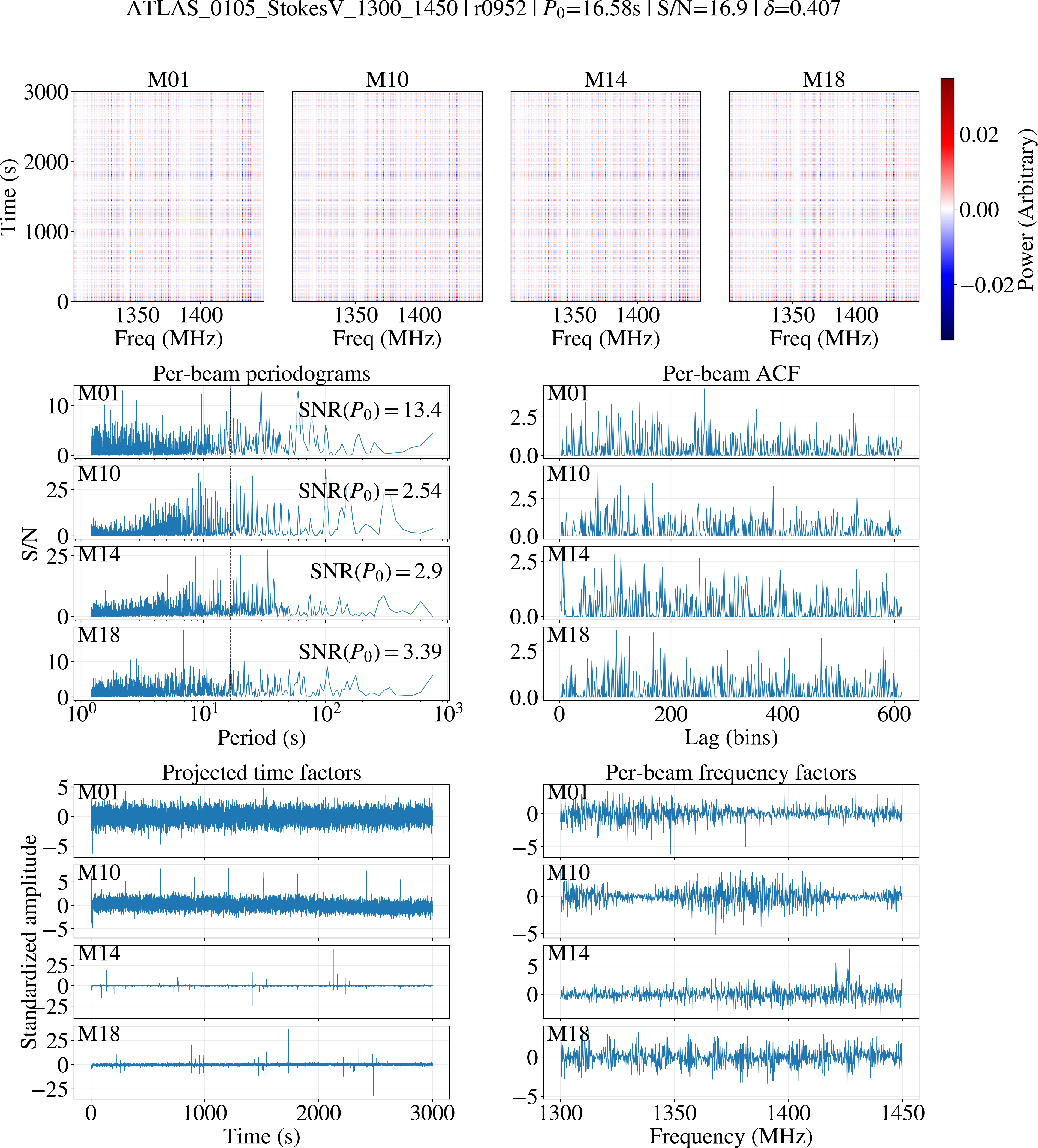}
  \caption{\label{fig:Cand_components}  Candidate component r55 with period of $P_0=18.87$ s in 1050--1140 MHz (a) and r952 with period of $P_0=16.58$ s 1300--1450 MHz (b) in our search. The layout of the two panels are the same as Figure~\ref{fig:simulate_signal}.}
\end{figure}

\begin{figure}[htbp]
    \centering
    \includegraphics[width=\textwidth]{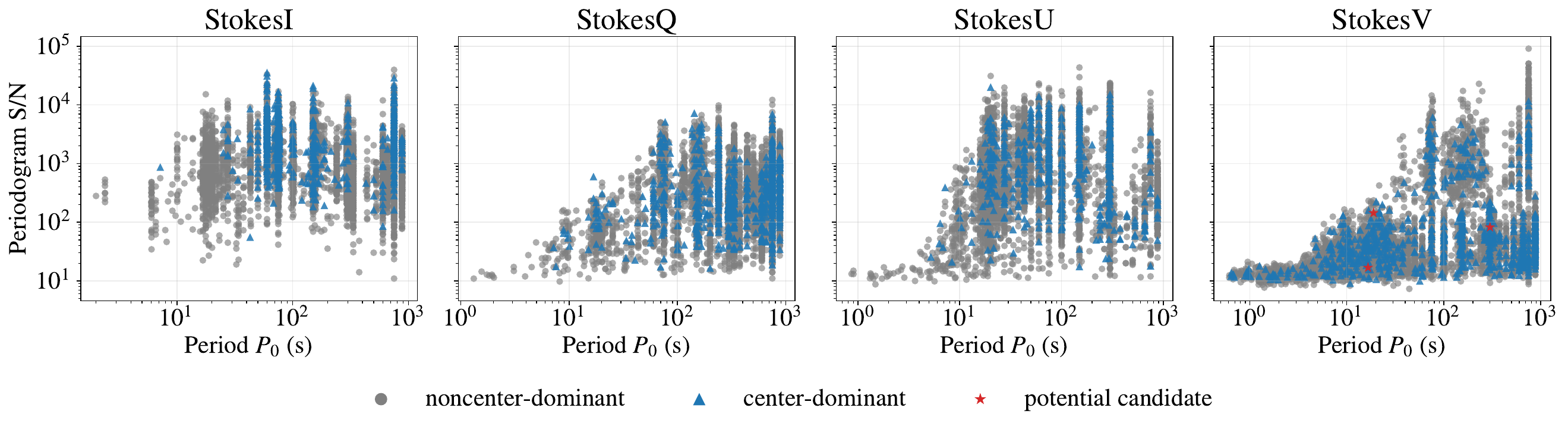}
    \caption{Distribution of components for Stokes $IQUV$ polarization channels in our observation. The central beam dominant components are denoted in blue triangle, and the three potential candidates are highlighted in red star.}
    \label{fig:period_snr_scatter}
\end{figure}

Figure~\ref{fig:period_snr_scatter} shows the distribution of all components in the $(P_0,\mathrm{SNR})$ plane for Stokes-$IQUV$. Across all polarization channels, the population is strongly skewed toward longer periods. Components with short periods are comparatively rare, while the density increases markedly from tens of seconds to several hundred seconds and beyond. A prominent feature is the occurrence of vertical banding and clustered structures, where many components share nearly the same period but span a wide range of periodogram SNR. This pattern suggests that the periods are not uniformly spread across the searched range. Instead, many components accumulate around a limited number of preferred timescales, forming vertical bands in the period-SNR plane. Within each vertical band, the periodogram SNR spans a wide dynamic range, indicating that the same characteristic timescale can appear with varying apparent significance. 

\section{Discussion} \label{sec:Discussion}
\subsection{Sensitivity}\label{subsec:Sensitivity}
The sensitivity of a radio observation can be determined by system equivalent flux density \citep{2013tra..book.....W,2017isra.book.....T}
\begin{equation}
  \mathrm{SEFD}=\frac{2k_\mathrm{B}T_\mathrm{sys}}{A_\mathrm{eff}},
  \label{SEFD}
\end{equation}
where $k_\mathrm{B}$ is the Boltzmann constant, $T_\mathrm{sys}$ is the system temperature, and $A_\mathrm{eff}$ is the effective collecting area. The sensitivity $A_\mathrm{eff}/T_\mathrm{sys}$ of FAST L-band 19 beam receiver is $\sim 2000 \, \mathrm{m^{2} \, K^{-1}}$ \citep{2011IJMPD..20..989N,2016RaSc...51.1060L,2019SCPMA..6259502J}. To detect a channel-width periodic signal, the minimum detectable flux density $S_{\min}$ can be calculated by \citep{ 2020MNRAS.497.4654M}
\begin{equation}
     S_{\min}\approx\frac{({\rm SNR})_{\min}\,{\rm SEFD}}
{\sqrt{n_{\rm pol}\,\Delta \nu_{\rm ch}\, T_{\rm obs}}}\sqrt{\frac{\delta}{1-\delta}},
\end{equation}
where $({\rm SNR})_{\min}=10$ is the signal-to-noise ratio threshold, $\Delta \nu_{\rm ch}\sim 0.122~{\rm MHz}$ is the frequency channel bandwidth, $n_\mathrm{pol}=2$ is the number of polarization channels adopted in the sensitivity estimate, $T_{\rm obs}=3000~{\rm s}$ is the observation duration, and $\delta$ is the duty cycle. Although the periodic search is performed for all four Stokes products $(I,Q,U,V)$, the EIRP limits reported here are computed as a Stokes-$I$ sensitivity estimate, therefore we adopt $n_{\rm pol}=2$, corresponding to the two independent orthogonal polarization channels used to form Stokes $I$. For the observation target at distance $d$, our minimum luminosity detection threshold can be quantified by equivalent isotropic radiated power (EIRP) by 
\begin{equation}
  \mathrm{EIRP}=4\pi d^2 S_{\min} \Delta \nu_{\rm ch}.
  \label{EIRP}
\end{equation}
The corresponding EIRP for each observation date is listed in the second column of Table \ref{table:EIRP_list}. The minimum detectable flux density $S_{\min}$  is calculated by the median $\delta$ value of Stokes-$I$ for each date.

\subsection{Figure of Merit for Signal Search}\label{subsec:FOM}
A commonly used figure of merit to evaluate the signal search is continuous-waveform transmitter figure of merit (CWTFM) defined in \cite{2017ApJ...849..104E}
\begin{equation}
     {\rm CWTFM}=\frac{\xi_{\rm ref} {\rm EIRP}}{N\left(\frac{\Delta \nu_{\rm tot}}{\nu_c}\right)},
     \label{CWTFM}
\end{equation}
where $N$ is the number of observations\footnote{In the original definition in \cite{2017ApJ...849..104E}, $N=N_{\rm stars}=n_{\rm stars}\times N_{\rm pointing}$ refers the total number of stars during the survey, with $N_{\rm pointing}$ being the number of pointings during the survey and $n_{\rm stars}$ being the number of stars per pointing. In our single-target observation, although $n_{\rm stars}$ is 1, we still carry out several times of tracking observations in different dates. Therefore, in this work, we refer $N$ to the number of observations. }, $\Delta \nu_{\rm tot}$ denotes the total frequency bandwidth, $\nu_c$ is the central frequency in our search and $\xi_{\rm ref}=10^{-10} \, {\rm W ^{-1}}$ is the reference normalization factor. 

To quantify the merits of our periodic signal search, we also adopt the periodic spectral signaltransmitter figure of merit (PSSTFM) defined in \cite{2023AJ....165..255S}
\begin{equation}
     {\rm PSSTFM}=\frac{\xi_{\rm ref} {\rm EIRP}}{N\left(\frac{\Delta \nu_{\rm tot}}{\nu_c}\right)\log_{10}\left( \frac{P_{\max}}{P_{\min}}\right)\log_{10}\left( \frac{\delta_{\max}}{\delta_{\min}}\right)},
     \label{PSSTFM}
\end{equation}

Compared with the CWTFM in Eq. (\ref{CWTFM}), PSSTFM in Eq. (\ref{PSSTFM}) includes additional terms involving period and duty cycle in the denominator. In this case, for a fixed CWTFM value, smaller PSSTFM indicates a more thorough search. The corresponding PSSTFM for each observation date are also listed in  Table \ref{table:EIRP_list}.

\begin{deluxetable}{lccccc}[htpb]
\tablecaption{EIRP, CWTFM, and PSSTFM values of our periodic signal search toward 3I/ATLAS.\label{table:EIRP_list}}
\tablehead{
\colhead{Observation Date} & 
\colhead{Distance (AU)} &
\colhead{Median $\delta$} &
\colhead{EIRP (W)} &
\colhead{CWTFM} &
\colhead{PSSTFM} 
}
\startdata
2025-10-29 & 2.3101 & 0.097 & 0.304 & $2.53\times10^{-11}$ & $4.72\times10^{-12}$ \\
2025-12-19 & 1.7977 & 0.190 & 0.272 & $2.27\times10^{-11}$ & $4.22\times10^{-12}$ \\
2026-01-05 & 1.9446 & 0.047 & 0.146 & $1.22\times10^{-11}$ & $2.27\times10^{-12}$
\enddata
\end{deluxetable}

\subsection{Interpretation of the CPD-based Periodicity Search Framework}

The CPD-based periodic search adopted in this work takes a tensor component as the basic analysis unit. Each component is characterized by a joint structure in time, frequency, and beam space, represented respectively by its temporal, frequency, and beam-domain factors. A periodic signal is not treated simply as a localized excess at frequency channels or within one beam, but as a separable multibeam dynamic-spectrum component with an associated periodicity, spectral morphology, and beam-distribution pattern. Such a representation is particularly well suited to the central-beam/reference-beam observing configuration adopted here, because it allows the center-dominant nature of the component to be evaluated directly through the CPD-based beam-domain diagnostics.

Periodic signals may appear in multiple distinct frequency channels. In such a situation, CPD provides a natural way to examine whether an individual periodic component remains concentrated in the central beam. The resulting analysis combines periodicity diagnostics directly with multibeam morphology, which is a central requirement for distinguishing center-dominant behavior from broadly distributed instrumental RFIs. Moreover, the components can be naturally ordered according to their energy contribution, providing a convenient hierarchy for visual inspection and statistical screening. Components with larger energy ratios generally represent the most prominent structures present in the data, while very low-energy components are more often consistent with weak residuals or noise-like fluctuations.

\section{Conclusion} \label{sec:Conclusion}

We perform channel-width periodic technosignature search toward 3I/ATLAS with FAST L-band multibeam receiver in the frequency range of 1.05-1.45 GHz. We propose an analysis framework based on the utilization of canonical polyadic decomposition (CPD) to multibeam data, so that we can split them into separated components with weighted rank and determine those components with center-dominant feature. Each component is characterized by beam dominance, periodicity, as well as ACF diagnostics in frequency. 

After using our CPD-based framework to the four Stokes products from the 3-hr 3I/ATLAS observation data, as well as visual inspeciton for each component, we report a nondetection of channel-width periodic signal within $P\in[0.7 {\rm s}, 1000{\rm s}]$ above our adopted periodogram-significance threshold of 10$\sigma$. This nondetection can place a constraint that there is no solid evidence for any periodic singal transmitter with EIRP above 0.146 W on 3I/ATLAS.

In the future, we plan to extend the periodic signal search with segment-consistency tests across the observing time, requiring candidate periodicities to persist coherently in independent time blocks rather than arising from localized fluctuations. This extension naturally enables the use of our psr backend data with microsecond-level sampling, thereby expanding the searchable period range to shorter timescales. We also plan to apply our CPD-based framework to multibeam targeted observations for periodic technosignature search with FAST. Signal with periodic modulation remains comparatively under-explored as a technosignature type relative to narrowband drifting searches, and incorporating it can help extend the coverage of plausible technosignature classes. 

\begin{acknowledgments}
We sincerely appreciate the staffs in FAST data center for approving our DDT application on FAST observing time. This work is supported by National Key R\&D Program of China (2023YFB4503305), the China Manned Space Program with grant No. CMS-CSST-2025-A01, the National Natural Science Foundation of China (Grants No12373109). This work made use of the data from FAST (Five-hundred-meter Aperture Spherical radio Telescope).  FAST is a Chinese national mega-science facility, operated by National Astronomical Observatories, Chinese Academy of Sciences.
\end{acknowledgments}



\appendix
\section{Computational Implementation of CPD}
\label{app:cpals}
The multibeam dynamic spectrum is represented as a dense third-order tensor $X\in\mathbb{R}^{T\times F\times B}$, with modes corresponding to time,
frequency, and beam index. The CP model is fitted using the standard alternating least-squares algorithm \citep[CP-ALS;][]{doi:10.1137/07070111X}. In each ALS iteration, two factor matrices are held fixed while the remaining factor matrix is updated by solving a linear least-squares euqation.  For the three modes in our tensor, the updates can be written as
\begin{equation}
\label{eq:update}
\begin{aligned}
A &\leftarrow
X_{(t)}(D\odot C)
\left[(C^{\rm T}C)\ast(D^{\rm T}D)+\eta I\right]^{-1},\\
C &\leftarrow
X_{(\nu)}(D\odot A)
\left[(A^{\rm T}A)\ast(D^{\rm T}D)+\eta I\right]^{-1},\\
D &\leftarrow
X_{(b)}(C\odot A)
\left[(A^{\rm T}A)\ast(C^{\rm T}C)+\eta I\right]^{-1},
\end{aligned}
\end{equation}
where $X_{(t)}$, $X_{(\nu)}$, and $X_{(b)}$ are the tensor unfoldings along the time, frequency, and beam modes, respectively. Here $\odot$ denotes the
Khatri--Rao product, $\ast$ denotes the Hadamard product, and $\eta I$ is a
small ridge-regularization term used for numerical stability.

We implement these updates directly in Python/NumPy by explicitly forming the tensor unfoldings and Khatri--Rao products and solving the regularized normal equations given in Equation~\eqref{eq:update}. After each full ALS iteration, the columns of $A$, $C$, and $D$ are normalized and the corresponding norms are absorbed into the component weights $\lambda_r$. This normalization fixes the arbitrary scaling among the three factor matrices, although the usual CPD sign and permutation indeterminacies remain. This implementation is algorithmically equivalent to the standard dense CP-ALS routines such as \texttt{TensorLy} \citep{tensorly} or \texttt{pyTensorlab} \citep{pytensorlab}, but it keeps direct control over the time--frequency--beam tensor construction,beam indexing, rank setting, and the subsequent periodogram and autocorrelation diagnostics.


\bibliography{sample701}{}
\bibliographystyle{aasjournalv7}


\end{CJK*}
\end{document}